%% file: main.tex
\documentclass[11pt]{article}
\usepackage[linesnumbered,vlined,boxed,ruled]{algorithm2e}

\usepackage{fullpage}
\usepackage{rpmacros}
\usepackage[T1]{fontenc}
\usepackage[usenames,dvipsnames]{color}
\usepackage{stmaryrd}
\usepackage{verbatim}
\RequirePackage[colorlinks=true]{hyperref}
\hypersetup{
	linkcolor=[rgb]{0.3,0.3,0.6},
	citecolor=[rgb]{0.2, 0.6, 0.2},
	urlcolor=[rgb]{0.6, 0.2, 0.2}
}
\usepackage{mathpazo}
\usepackage{bm}
\usepackage{todonotes}
\usepackage{lipsum}
\usepackage{setspace}
\usepackage{scrextend}

\input{thmmacros}

\input{lazymacros} 
\input{customurlbst/bibmacros}

\makeatletter
\newcommand{\RemoveAlgoNumber}{\renewcommand{\fnum@algocf}{\AlCapSty{\AlCapFnt\algorithmcfname}}}
\newcommand{\RevertAlgoNumber}{\algocf@resetfnum}
\makeatother

{\expandafter\comment}%
{\expandafter\endcomment}

\onehalfspace

\title{Constructing Faithful Homomorphisms over Fields of Finite Characteristic}
\author{
	Prerona Chatterjee\thanks{Blavatnik Scool of Computer Science, Tel Aviv University, Israel. This work was done while the author was a PhD student in TIFR, Mumbai; and was supported by a fellowship of the DAE, India. Email: prerona.ch@gmail.com}
	\and
	Ramprasad Saptharishi\thanks{Tata Institute of Fundamental Research, Mumbai, India. Research supported by Ramanujan Fellowship of DST. Email: ramprasad@tifr.res.in}
}

\begin{document}
	
	\maketitle

	\begin{abstract}
		We study the question of algebraic rank or transcendence degree preserving homomorphisms over finite fields. This concept was first introduced by Beecken, Mittmann and Saxena~\cite{BMS13}, and exploited by them, and Agrawal, Saha, Saptharishi and Saxena~\cite{ASSS16} to design algebraic independence based identity tests using the Jacobian criterion over characteristic zero fields. An analogue of such constructions over finite characteristic fields was unknown due to the failure of the Jacobian criterion over finite characteristic fields. 
		
		Building on a recent criterion of Pandey, Saxena and Sinhababu~\cite{PSS18}, we construct explicit faithful maps for some natural classes of polynomials in the positive characteristic field setting, when a certain parameter called the \emph{inseparable degree} of the underlying polynomials is bounded (this parameter is always $1$ in fields of characteristic zero). This presents the first generalisation of some of the results of Beecken \etal~\cite{BMS13} and Agrawal \etal~\cite{ASSS16} in the positive characteristic setting.		
	\end{abstract}
        
\newpage
\section{Introduction}
	
Multivariate polynomials are fundamental objects in mathematics. 
These are the primary objects of study in algebraic complexity with regard to classifying their hardness as well as algorithmic tasks involving them. 
The standard computational model for computing multivariate polynomials is \emph{algebraic circuits}. 
These are directed acyclic graphs with internal nodes labelled by `$+$' and `$\times$' gates having the obvious operational semantics, and  leaves are labelled by the input variables or field constants. 
	
An important concept about relationships between polynomials is the notion of \emph{algebraic dependence}. 
A set of polynomials $\vecf = \set{f_1,\ldots, f_m} \subset \F[\vecx]$ is said to be \emph{algebraically dependent} if and only if there is some nonzero polynomial combination of $\set{f_1,\ldots, f_m}$ that is zero. 
Such a nonzero polynomial $A(z_1,\ldots, z_m) \in \F[\vecz]$, if one exist, for which $A(f_1,\ldots, f_m) = 0$ is called the \emph{annihilating polynomial} for the set $\set{f_1, \ldots, f_m}$. 
For instance, if $f_1 = x$, $f_2 = y$ and $f_3 = x^2+y^2$, then $A = z^2_1 + z^2_2 - z_3$ is an annihilator. Note that the underlying field is very important. 
For example, the polynomials $x+y$ and $x^p + y^p$ are algebraically dependent over $\F_p$, but algebraically independent over a characteristic zero field.
	
Algebraic independence is very well-studied and it is known that algebraically independent subsets of a given set of polynomials form a \emph{matroid} (see \cite{O92}). 
Hence, the size of the maximum algebraically independent subset of $\vecf$ is well-defined and is called the \emph{algebraic rank} or \emph{transcendence degree} of $\vecf$. 
We denote it by $\algrank(\vecf) = \algrank(f_1, \ldots, f_m)$.
	
Several computational questions arise from the above definition. 
For instance, given a set of polynomials $\vecf = \set{f_1, \ldots, f_m}$, each $f_i$ given in its dense representation, can we compute the algebraic rank of this set efficiently?
What if the $f_i$'s are provided as algebraic circuits?
Such a nonzero polynomial $A(z_1,\ldots, z_m) \in \F[\vecz]$, if one exist, for which $A(f_1,\ldots, f_m) = 0$ is called the \emph{annihilating polynomial} for the set $\set{f_1, \ldots, f_m}$. 
Furthermore, in instances when $\algrank(\vecf) = m-1$, Kayal~\cite{K09} showed that the smallest degree annihilating polynomial is unique. 
There could be various questions about the minimal degree annihilator in this case. For instance, can we compute it efficiently? 
Kayal~\cite{K09} showed that even checking if the constant term of the annihilator is zero is $\NP$-hard, and evaluating the annihilator at a given point is $\#\P$-hard. 
In fact, recently Guo, Saxena, Sinhababu \cite{GSS19} showed that even in the general case, checking if the constant term of every annihilator is zero is $\NP$-hard. 
This effectively rules out any attempt to compute the algebraic rank via directly checking properties of the annihilating polynomials.
	
Despite this, over fields of characteristic zero, algebraic rank has an alternate characterisation via the Jacobian criterion. 
Jacobi \cite{J41} showed that the algebraic rank of a set of polynomials $\vecf (\subseteq \F[\vecx])$ is given by the linear rank (over the rational function field $\F(\vecx)$) of the Jacobian of these polynomials. 
This immediately yields a randomized polynomial time algorithm to compute the algebraic rank of a given set of polynomials by computing the rank of the Jacobian evaluated at a random point due to the polynomial identity lemma \cite{O22,S80,Z79,DL78}.
	
\subsubsection*{Faithful homomorphisms and PIT}
Algebraic independence shares a lot of similarities with linear independence due to the matroid structure. 
One natural task is to find a \emph{rank-preserving transformation} in this setting. This is defined by what are called \emph{faithful homomorphisms}.
	
\begin{definition}[Faithful homomorphisms \cite{BMS13}] \label{defn:faithful}
	Let $\vecf = \set{f_1,\ldots, f_m} \subseteq \F[\vecx]$ be a set of polynomials. If $\mathbb{K}$ is an extension field of $\F$, a homomorphism 
	$\Phi: \F[\vecx] \rightarrow \mathbb{K}[\vecy]$ 
	is said to be an \emph{$\F$-faithful} homomorphism for $\set{f_1,\ldots, f_m}$ if
	\[
		\algrank_\F \set{f_1,\ldots, f_m} = \algrank_\F \set{\Phi(f_1),\ldots, \Phi(f_m)}.\qedhere
	\]
\end{definition}
	
Ideally, we would like a faithful homomorphism with $\abs{\vecy} \approx \algrank \set{\vecf}$ and $\mathbb{K} = \F$. 
Beecken, Mittmann and Saxena \cite{BMS13} showed that a \emph{generic} $\F$-linear homomorphism to $\algrank(\vecf)$ many variables would be an $\F$-faithful homomorphism with high probability. 
	
One important consequence of faithful homomorphisms is that they preserve nonzeroness of any polynomial composition of $f_1, \ldots, f_m$.
	
\begin{lemma}[\cite{BMS13,ASSS16}]\label{lem: connect}
	Suppose $f_1, \ldots, f_m \in \F[x_1, \ldots, x_n]$ and $\Phi$ is an $\F$-faithful homomorphism for $\set{f_1, \ldots, f_m}$. 
	Then, for any circuit $C(z_1, \ldots, z_m) \in \F[z_1,\ldots, z_m]$, we have \[C(f_1, \ldots, f_m) = 0 \Leftrightarrow C(\Phi(f_1),\ldots, \Phi(f_m)) = 0. \]
\end{lemma}

Thus, constructing explicit faithful homomorphisms can also be used for polynomial identity testing (PIT), which is the task of checking if a given algebraic circuit $C$ computes the identically zero polynomial. 
For PIT, the goal is to design a deterministic algorithm that runs in time polynomial in the size of the circuit. 
There are two types of PIT algorithms, \emph{whitebox} and \emph{blackbox} --- in the blackbox setting, we are only provided evaluation access to the circuit and some of its parameters (such as degree, number of variables, size etc.). 
Thus blackbox PIT algorithms for a class $\ckt$ is equivalent to constructing a \emph{hitting set}, which is a small list of points in $S \subset \F^n$ such that any nonzero polynomial $f\in \ckt$ is guaranteed to evaluate to a nonzero value on some $\veca \in S$.
	
It follows from \autoref{lem: connect} that if we can construct explicit $\F$-faithful homomorphisms for a set $\set{f_1,\ldots, f_m}$ whose algebraic rank is $k \ll n$, then we have a \emph{variable reduction} that preserves the nonzeroness of any composition $C(f_1,\ldots, f_m)$. 
This approach was used by Beecken, Mittmann and Saxena \cite{BMS13} and Agrawal, Saha, Saptharishi, Saxena~\cite{ASSS16}, in the characteristic zero setting, to design identity tests for several subclasses by constructing faithful maps for $\set{f_1,\ldots, f_m}$ with algebraic rank at most $k = O(1)$, when
\begin{itemize}
	\item each $f_i$ is a sparse polynomial,
	\item each $f_i$ is a product of multilinear, variable disjoint, sparse polynomials,
	\item each $f_i$ is a product of linear polynomials,
\end{itemize}
and further generalisations.
	
All the above constructions crucially depend on the fact that the rank of the Jacobian captures algebraic independence. 
However, this fact is true only over fields of characteristic zero and hence all the above results no longer hold over fields of positive characteristic. 
	
\subsubsection*{Algebraic independence over finite characteristic}
	
A standard example to exhibit the failure of the Jacobian criterion over fields of finite characteristic, is $\set{x^{p-1}y, y^{p-1}x}$ --- these polynomials are algebraically independent over $\F_p$ but the Jacobian is \emph{not} full-rank over $\F_p$. 
Pandey, Saxena and Sinhababu~\cite{PSS18} characterised the extent of failure of he Jacobian criterion for $\set{f_1,\ldots, f_m}$ by a notion called the \emph{inseparable degree} associated with this set (formally defined in \autoref{sec:field-theory}). 
Over characteristic zero fields, this is always $1$ but over fields of characteristic $p$ this is a power of $p$. 
In their work, Pandey \etal{} presented a Jacobian-like criterion to capture algebraic independence.
Informally, each row of the \emph{generalized Jacobian matrix} is obtained by taking the Taylor expansion of $f_i(\vecx + \vecz)$ about a generic point, and truncating to just the terms of degree up to the \emph{inseparable degree}\footnote{Over characteristic zero, the inseparable degree is $1$ and this is just the vector of first order partial derivatives}. 
The exact characterisation is more involved and is presented in \autoref{sec:PSS} but we just state their theorem here. 
	
\begin{restatable}{theorem}{PSS}\emph{\cite{PSS18}}\label{thm:criterion}
	Let $\set{f_1, \ldots, f_k}$ be a set of $n$-variate polynomials over a field $\F$ with inseparable degree $t$. Also, for a generic point $\vecz$, let $\Ech_t(f_i) = \deg_{\leq t}(f_i(\vecx + \vecz) - f_i(\vecz))$. 
	Then, they are algebraically dependent if and only if 
	\[
		\exists (\alpha_1, \ldots, \alpha_k) (\neq \mathbf{0}) \in \F(\vecz)^k \text{ s.t. } \sum_{i=1}^k \alpha_i \cdot \Ech_t(f_i) = 0 \mod \inangle{\Ech_t(f_1), \ldots, \Ech_t(f_k)}^{\geq 2}_{\F(\vecz)} + \inangle{\vecx}^{t+1}.
	\]
\end{restatable}

We note that although the statement above seems slightly different from the one in \cite{PSS18}, it is not too hard to see that they are actually equivalent. 
In their paper, Pandey \etal{} have stated their criterion in terms of functional dependence. 
However, stated this way, it clearly generalises the traditional Jacobian criterion. 
	
In the setting when the \emph{inseparable degree} is constant, this characterisation yields a randomized polynomial time algorithm to compute the algebraic rank. 
Thus, a natural question is whether this criterion can be used to construct faithful homomorphisms for similar classes of polynomials as studied by Beecken \etal~\cite{BMS13} and Agrawal \etal~\cite{ASSS16}. 
	
\begin{remark}
	Recently, Guo \etal~\cite{GSS19} showed that the task of testing algebraic independence is in $\AM \cap \coAM$ via a very different approach. 
	However, it is unclear if their algorithm also yields constructions of faithful homomorphisms or applications to PIT in restricted settings. 
\end{remark}
	
Following up on the criterion of Pandey, Saxena and Sinhababu~\cite{PSS18} for algebraic independence over finite characteristic, we extend the results of Beecken \etal~\cite{BMS13} and Agrawal \etal~\cite{ASSS16} to construct faithful homomorphisms for some restricted settings. 
	
\begin{restatable}{theorem}{FaithfulMaps}\label{thm:FaithfulConstruct}
	Let $f_1,\ldots, f_m \in \F[x_1,\ldots, x_n]$ be such that $\algrank \set{f_1,\ldots, f_m} = k$ and the inseparable degree is $t$. 
	If $t$ and $k$ are bounded by a constant, then we can construct a polynomial (in the input length) sized list of homomorphisms of the form $\Phi:\F[\vecx] \rightarrow \F(s)[y_0, y_1,\ldots, y_k]$ such that at least one of them is guaranteed to be $\F$-faithful for the set $\set{f_1,\ldots, f_m}$, in the following two settings:
	\begin{itemize}
		\item When each of the $f_i$'s are sparse polynomials,
		\item When each of the $f_i$'s are products of variable disjoint, multilinear, sparse polynomials.
	\end{itemize}
\end{restatable}
	
Prior to this, construction of faithful homomorphisms over finite fields was known only in the setting when each $f_i$ has small individual degree \cite{BMS13}. 
Over characteristic zero fields, the inseparable degree is always $1$ and hence the faithful maps constructed in \cite{BMS13}, \cite{ASSS16} over such fields can be viewed as special cases of our constructions. 
	
The above theorem also holds for a few other models studied by Agrawal \etal~\cite{ASSS16} (for instance, occur-$k$ products of sparse polynomials).
We mention the above two models just as an illustration of lifting the recipe for faithful maps from \cite{BMS13,ASSS16} to the finite characteristic setting. 
As corollaries, we get efficient PIT algorithms for these models.
	
\begin{restatable}{corollary}{CorSparsePIT}
	If $\set{f_1, \ldots, f_m} \in \F[x_1, \ldots, x_n]$ is a set of $s'$-sparse polynomials with algebraic rank $k$ and inseparable degree $t$ where $k,t = O(1)$. 
	Then, for the class of polynomials of the form $C(f_1,\ldots, f_m)$ for any polynomial $C(z_1,\ldots, z_m) \in \F[\vecz]$, there is an explicit hitting set of size $\inparen{s'\cdot \deg(C)}^{O(1)}$.
\end{restatable}
	
\begin{restatable}{corollary}{CorMLPIT}	\label{cor:PIT-multilinear}
	Let $\ckt = \sum_{i=1}^{m} T_i$ be a depth-$4$ multilinear circuit of size $s$, where each $T_i$ is a product of variable-disjoint, $s$-sparse polynomials. 
	Suppose $\set{T_1,\ldots, T_m} \in \F[x_1, \ldots, x_n]$ is a set of polynomials with algebraic rank $k$ and inseparable degree $t$ where $k,t = O(1)$. 
	Then, for the class of polynomials of the form $C(T_1,\ldots, T_m)$ for any polynomial $C(z_1,\ldots, z_m) \in \F[\vecz]$, there is an explicit hitting set of size $\inparen{s\cdot \deg(C)}^{O(1)}$.
\end{restatable}
	
\subsubsection*{Comparison with the PIT of \cite{PSS18}}
	
Pandey \etal \cite{PSS18} also give a PIT result in their work for circuits of the form $\sum_i  \inparen{f_{i,1} \cdots f_{i,m}}$ where $\algrank \set{f_{i,1}, \ldots, f_{i,m}} \leq k$ for every $i$ and each $f_{i,j}$ is a degree $d$ polynomial in $\F[x_1, \ldots, x_n]$.
They extend the result of Kumar and Saraf \cite{KS17} to arbitrary fields by giving quasi-polynomial time hitting sets if $kd$ is at most poly-logarithmically large.
	
\autoref{cor:PIT-multilinear} however is incomparable to the PIT of Pandey \etal~\cite{PSS18} for the following reasons:
\begin{itemize}
	\item The algebraic rank bound in the case of \cite{PSS18,KS17} is a gate-wise bound rather than a global bound. 
	Thus, in principle, it could be the case that $\algrank\set{f_{i,1},\ldots, f_{i,m}}$ is bounded by $k$ for each $i$ but this would not necessarily translate to a bound on $\algrank\setdef{\prod_j f_{i,j}}{i}$ as demanded in \autoref{cor:PIT-multilinear}. 
	Hence, in this regard, the PIT of \cite{PSS18,KS17} is stronger.
	\item In the regime when we have $\algrank\setdef{\prod_j f_{i,j}}{i}$ and the inseparable degree of this set to be bounded by a constant, \autoref{cor:PIT-multilinear} presents an explicit hitting set of polynomial size, whereas it is unclear if \cite{PSS18,KS17} provide any non-trivial upper bound as this does not translate to any bound on $\algrank\set{f_{i,1},\ldots, f_{i,m}}$. 
\end{itemize}
	
\subsubsection*{On other models studied by Agrawal \etal~\cite{ASSS16}}
	
Our results, in its current form, do not extend directly some of the other models studied by Agrawal \etal~\cite{ASSS16}, most notably larger depth multilinear formulas. 
The primary hurdle appears to be the \emph{recursive} use of explicit faithful homomorphisms for larger depth formulas. 
In the characteristic $p$ setting, unfortunately, it is unclear if a bound on the inseparable degree of the original gates can be used to obtain a bound on the inseparable degree of other sets of polynomials considered in the recursive construction of Agrawal \etal~\cite{ASSS16}. 
	
\subsection{Proof overview}
	
The general structure of the proof follows the outline of Agrawal \etal~\cite{ASSS16}'s construction of faithful homomorphisms in the characteristic zero setting. Roughly speaking, this can be described in the following steps:
	
\begin{description}
	\item[Step 1]: For a \emph{generic linear map} $\Phi: \vecx \rightarrow \F(s)[y_1,\ldots, y_k]$, write the Jacobian of the set of polynomials $\set{f_1 \circ \Phi, \cdots, f_k\circ \Phi}$. 
	Thus can be described succinctly as a matrix product of the form
	\[
		J_\vecy(f \circ \Phi) = \Phi(J_\vecx(\vecf)) \cdot  J_\vecy(\Phi(\vecx)). 
	\]
	\item[Step 2]: We know that $J_\vecx(\vecf)$ is full rank. 
	Ensure that $\Phi(J_\vecx(\vecf))$ (where $\Phi$ is applied to every entry of the matrix $J_\vecx(\vecf)$) remains full rank. 
	This can be done if $\vecf$'s are some structured polynomials such as sparse polynomials, or variable-disjoint products of sparse polynomials etc.
	\item[Step 3]: Choose the map $\Phi$ so as to ensure that
	\[
		\rank(\Phi(J_\vecx(\vecf)) \cdot  J_\vecy(\Phi(\vecx))) = \rank(\Phi(J_\vecx(\vecf))).
	\]
	This is typically achieved by choosing $\Phi$ so as to make $J_\vecy(\Phi(\vecx))$ a \emph{rank-extractor}. 
	It was shown by Gabizon and Raz~\cite{GR08} that a parametrized Vandermonde matrix has this property and this allows us to work with a homomorphism of the form (loosely speaking)
	\[
		\Phi: x_i \mapsto \sum_{j=1}^k s^{ij} y_j. 
	\]
\end{description}
	
We would like to execute essentially the same sketch over fields of finite characteristic but we encounter some immediate difficulties. 
The criterion of Pandey \etal~\cite{PSS18} over finite characteristic is more involved but it is reasonably straightforward to execute Steps 1 and 2 in the above sketch using the chain rule of (Hasse) derivatives. 
The primary issue is in executing Step 3 and this is for two very different reasons.
	
The first is that, unlike in the characteristic zero setting, the analogue of the matrix $J_\vecy(\Phi(\vecx))$ has many correlated entries. 
In the characteristic zero setting, we have complete freedom to choose $\Phi$ so that $J_\vecy(\Phi(\vecx))$ can be any matrix that we want. 
Roughly speaking, we only have $n \cdot k$ parameters to define $\Phi$ but the analogue of $J_\vecy(\Phi(\vecx))$ is much larger in the finite characteristic setting. 
Fortunately, there is just about enough structure in the matrix that we can show that it continues to have some rank-preserving  properties. 
This is done in \autoref{sec:rank-conc}.
	
The second hurdle comes from the subspace that we need to work with in the modified criterion. 
The \emph{rank-extractor} is essentially parametrized by the variable $s$. In order to show that it preserves the rank of $\Phi(J_\vecx(\vecf))$ under right multiplication, we would like to ensure that the variable $s$ effectively does not appear in this matrix. 
In the characteristic zero setting, this is done by a suitable restriction on the other variables to remove any dependencies on $s$ in $\Phi(J_\vecx(\vecf))$. 
Unfortunately, in the criterion of Pandey \etal~\cite{PSS18}, we have to work modulo some suitable subspace and these elements introduce other dependencies on $s$ that appear to be hard to remove. 
Due to this hurdle, we are unable to construct $\F(s)$-faithful homomorphisms even in restricted settings. 
	
However, we observe that for the PIT applications, we are merely required to ensure that $\set{f_1 \circ \Phi, \ldots, f_k\circ \Phi}$ remain $\F$-algebraically independent instead of $\F(s)$-algebraically independent. 
With this weaker requirement, we can obtain a little more structure in the subspace involved and that lets us effectively execute Step 3. 
	
\subsubsection*{Structure of the paper}
	
We begin with a description of some preliminaries that are necessary to understand the criterion of Pandey, Saxena and Sinhababu~\cite{PSS18} in the next section. 
Following that, in \autoref{sec:rank-conc}, we show that certain Vandermonde-like matrices have \emph{rank-preserving properties}. 
We use these matrices to give a recipe of constructing faithful maps, in \autoref{sec:recipe}, and execute this for the settings of \autoref{thm:FaithfulConstruct} in \autoref{sec:execution}.

\section{Preliminaries}

\subsection{Notations}
	
\begin{itemize}
	\item For a positive integer $m$, we will use $[m]$ to denote set $\set{1,2,\ldots, m}$. 
	\item We will use bold face letters such as $\vecx$ to denote a set of indexed variables $\set{x_1,\ldots, x_n}$. 
	In most cases the size of this set would be clear from context.
	Extending this notation, we will use $\vecx^{\vece}$ to denote the monomial $x_1^{e_1}\cdots x_n^{e_n}$. 
	\item For a set of polynomials $f_1,\ldots, f_m$, we will denote by $\inangle{f_1,\ldots, f_m}_\mathbb{K}$ the set of all $\mathbb{K}$-linear combinations of $f_1,\ldots, f_m$. 
	Extending this notation, we will use $\inangle{f_1,\ldots, f_m}_\mathbb{K}^{r}$ to denote the set of all $\mathbb{K}$-linear combinations of $r$-products $f_{i_1}\cdots f_{i_r}$ (with $i_1,\ldots, i_r\in [m]$) and $\inangle{f_1,\ldots, f_m}_\mathbb{K}^{\geq r}$ similarly. 
	In instances when we just use $\inangle{f_1,\ldots, f_m}$, we will denote the \emph{ideal} generated by $f_1,\ldots, f_m$. 
\end{itemize}
	
\subsection{Hitting set generators}

Hitting set generators are defined as follows.
	
\begin{definition}[Hitting set generators (HSG)]
	Let $\ckt$ be a class of $n$-variate polynomials. 
	A tuple of polynomials $\mathcal{G} = (G_1(\alpha),\ldots, G_n(\alpha))$ is a \emph{hitting set generator} for $\ckt$ if for every nonzero polynomial $P(\vecx) \in \ckt$ we have $P(G_1(\alpha),\ldots, G_n(\alpha))$ is a nonzero polynomial in $\alpha$.
		
	The degree of this generator is defined to be $\max \deg(G_i)$. 
\end{definition}
	
Intuitively, such a tuple can be used to \emph{generate} a hitting set for $\ckt$ by running over several instantiations of $\alpha$. 
Also, it is well known that any hitting set can be transformed into an HSG via interpolation. 
	
\subsection{Isolating weight assignments}
Suppose $\wt : \set{x_i} \to \N$ is a weight assignment for the variables $\set{x_1, \ldots, x_n}$.
We can extend it to define the weight of a monomial as follows.
\[
	\wt(\vecx^\vece) = \sum_{i=1}^{n} e_i \cdot \wt(x_i)
\]

\begin{definition}
	A weight assignment $\wt : \set{x_i} \to \N$ is said to be isolating for a set $S$ of monomials if every pair of distinct monomials in $S$ receives distinct weights.
\end{definition}
	
Note that if the highest degree of a monomial in $S$ is $d$, then assigning the weight $\wt(x_i) = (d+1)^i$ is trivially isolating for $S$. 
However, in this case the weight of a monomial can become exponentially large in $n$. 
	
In the case when $\abs{S} = \poly(n)$, results by Klivans and Spielman~\cite{KS01} or Agrawal and Biswas~\cite{AB03} show that if we define $\wt(x_i) = (d+1)^i \mod p$, then it suffices to go over $\poly(n)$ many `$p$'s to guarantee that one of these weight assignments isolates the monomials in $S$. 
The weight of a monomial in this case is thus bounded by $\poly(n)$.
	
\subsection{Some field theoretic preliminaries}	\label{sec:field-theory}
We present some basic preliminaries about field extensions. 
	
\begin{definition}
	A polynomial is said to be separable if it does not have repeated roots in a field where it factorises completely.
\end{definition}
	
Over characteristic zero fields, ever irreducible univariate polynomial is separable since it cannot have a common root with its derivative. 
However, this is not the case over fields of finite characteristic as derivatives of non-trivial polynomials could become zero. 
This adds some subtlety in field extensions over finite characteristic.
	
We mention some basic facts about field extensions; these may be found in any standard text for field theory \cite{I94}. 
\begin{enumerate}
	\item An extension $\mathbb{K}/\F$ is said to be algebraic if every element in $\mathbb{K}$ is the root of some polynomial over $\F$. Otherwise, it is transcendental. 
	\item For a transcendental extension $\mathbb{K}/\F$, a transcendence basis is a maximal subset of $\mathbb{K}$ that is algebraically independent over $\F$. 
	An extension $\mathbb{K}/\F$ is purely transcendental if there is a transcendence base $S \subseteq \F$ such that $\mathbb{K} = \F(S)$.
	\item An algebraic extension $\mathbb{K}/\F$ is said to be separable if the minimal polynomial of every element in $\mathbb{K}$ is separable.
		
	An example of an algebraic extension that is \emph{not} separable is $\F_p(x) / \F_p(x^p)$. 
	The minimal polynomial $\mu(z)$ for $x$ over $\F_p(x^p)$ is $z^p - x^p$, which is not separable. 
		
	Further, if $\mathbb{K} = \F(\alpha_1, \ldots, \alpha_n)$ is an algebraic extension of $\F$, then $\mathbb{K}/ \F$ is separable if and only if the minimal polynomials of $\alpha_i$ over $\F$ is separable for each $i$. 
\end{enumerate}
	
For an algebraic extension $\mathbb{K}/\F$ over characteristic $p$ the \emph{separable closure} of $\F$ in $\mathbb{K}$, denoted by $\Sep(\mathbb{K}/\F)$, is defined as
\[
	\Sep(\mathbb{K}/\F) = \setdef{\alpha \in \mathbb{K}}{\text{the minimal polynomial of } \alpha \text{ is separable over $\F$}}. 
\]

For every element $\alpha$ in $\mathbb{K} \setminus \Sep(\mathbb{K}/\F)$, we would have that $\alpha^{p^i} \in \Sep(\mathbb{K}/\F)$ for some positive integer $i$. 
Thus, the extension $\mathbb{K}/\F$ splits into two extensions $\mathbb{K} \geq \Sep(\mathbb{K}/\F) \geq \F$ where the latter is a \emph{separable algebraic} extension and the former is a \emph{purely inseparable} algebraic extension. 
	
\begin{definition}[Inseparable degree of algebraic extensions]
	For an algebraic extension $\mathbb{K}/\F$ of characteristic $p$, the \emph{inseparable degree} of the extension, denoted by $\insepdeg(\mathbb{K}/\F)$, is the smallest $t$ such that $x^{t} \in \Sep(\mathbb{K}/\F)$ for every $x\in \mathbb{K}$. 
\end{definition}
	
\begin{remark}
	The above definition deviates slightly from the standard definition texts on field theory, where the \emph{inseparable degree} is defined to be the degree of the extension $\mathbb{K}/\Sep(\mathbb{K}/F)$. 
	The definition above is the one used by Pandey, Saxena and Sinhababu~\cite{PSS18} in their criterion and we stick with it in this paper. 
\end{remark}
	
We would like to extend this definition to non-algebraic extensions. 
Let $\set{f_1,\ldots, f_m}$ be a set of polynomials over $\F$. 
We will be interested in the extension $\F(\vecx) = \F(x_1,\ldots, x_n)$ over $\F(f_1,\ldots, f_m)$. 
Suppose $\set{f_1,\ldots, f_k}$ is a separable transcendence basis of $\set{f_1,\ldots, f_m}$. 
Using the matroid property of algebraically independent polynomials, there exists $x_{i_{k+1}},\ldots, x_{i_n}$ such that $\set{f_1,\ldots,f_k, x_{i_{k+1}},\ldots, x_{i_n}}$ is algebraically independent as well. 
Now, since $\F(\vecx)$ is algebraic over $\F(f_1,\ldots, f_k,x_{i_{k+1}},\ldots, x_{i_n})$, we can talk about the inseparable degree of this algebraic extension. 
We use this to define a suitable notation of inseparable degree\footnote{This definition is non-standard, but is sufficient for the purposes of this paper and the criterion of Pandey, Saxena and Sinhababu~\cite{PSS18}} for a set of algebraically independent polynomials.  
	
\begin{definition}[Inseparable degree of a set of polynomials]
	Let $\vecf = \set{f_1,\dots, f_m}$ be a set of polynomials over a field $\F$ of characteristic $p$. 
	For a set $S \subseteq [n]$, define $\vecx_S = \setdef{x_i}{i \in S}$. We shall define $\insepdeg(\set{f_1,\ldots, f_k})$ to be
	\[
		\min \setdef{\insepdeg\inparen{\F(\vecx)/\F(\vecf, \vecx_S)}}
		{\begin{array}{c}\text{$|S| = n - \algrank(\vecf)$ and}\\ 
			\text{$\F(\vecf,\vecx_S)/\F(\vecf)$ is purely transcendental}
		\end{array}}
	\]
\end{definition}
	
Intuitively, every extension can be thought of as purely transcendental, followed by a separable algebraic, followed by a purely inseparable algebraic extension. 
The above definition used the inseparable degree of the purely inseparable part of this in the general case. 
	
With this background, we are now ready to state the criterion for algebraic independence over fields of finite characteristic. 
Similar to the Jacobian Criterion, Pandey, Saxena and Sinhababu~\cite{PSS18} reduce the problem of checking algebraic independence to that of checking linear independence. 
However, their criterion is slightly more subtle in the sense that we will have to check the linear independence of a set of vectors modulo a large subspace.
	
\subsection{The PSS Criterion over fields of finite characteristic}\label{sec:PSS}
	
A set of polynomials $\set{f_1, \ldots, f_m} \in \F[x_1, \ldots, x_n]$ is said to be algebraically dependent if there exists a polynomial $0 \neq A \in \F[z_1, \ldots, z_m]$ such that $A(f_1, \ldots, f_m) = 0$. 
If such a polynomial $A(\vecz)$ exists, we call it the annihilating polynomial for $\set{f_1, \ldots, f_m}$.
	
However given a set of polynomials $\vecf = \set{f_1, \ldots, f_m} \in \F[\vecx]$, finding the annihilating polynomial if one exists is \emph{hard}~\cite{K09,GSS19}. 
Nevertheless if the underlying field $\F$ has characteristic zero, the Jacobian Criterion \cite{J41} reduces the question of checking whether a given set of polynomials is algebraically dependent to the question of checking whether a corresponding set of vectors is linearly dependent.
	
\subsubsection*{The Jacobian Criterion}
For $f_1, \ldots, f_m \in \F[x_1, \ldots, x_n]$, the Jacobian matrix is defined as 
\[
	\J_\vecx(\vecf) = 
	\begin{bmatrix}
		\partial_{x_1}(f_1) & \partial_{x_1}(f_2) & \ldots & \partial_{x_1}(f_m)\\
		\partial_{x_2}(f_1) & \partial_{x_2}(f_2) & \ldots & \partial_{x_2}(f_m)\\
		\vdots & \vdots & \ddots & \vdots\\
		\partial_{x_n}(f_1) & \partial_{x_n}(f_2) & \ldots & \partial_{x_n}(f_m)
	\end{bmatrix}
\]
With this definition, the Jacobian criterion \cite{J41} is as follows.
	
\begin{theoremwp}[Jacobian criterion]\label{thm:JacCriterion}
	If $\F$ is a field of characteristic zero, then $f_1, \ldots, f_m \in \F[\vecx]$ are algebraically independent if and only if $\J_\vecx(\vecf)$ has full rank over the rational function field $\F(\vecx)$.
\end{theoremwp}
	
As mentioned earlier,  this criterion is not true over fields that have finite characteristic. 
For $f_1 = x^{p-1}y$ and $f_2 = xy^{p-1}$, if the underlying field is $\F_p$, then $\det(\J(f_1, f_2)) = 0$ even though they are algebraically independent. 
The key insight of Pandey \etal~\cite{PSS18} is to observe that the rows of the Jacobian matrix, which are first order partial derivatives, are the linear terms present in the Taylor expansion of $f(\vecx)$ around a generic point $\vecz$. 
Generalising this, they study higher order terms of the Taylor expansion around a generic point to come up with a modified criterion that works over all fields. 
	
\subsubsection*{Taylor Expansion and Hasse Derivatives}
	
Define the following operator $\Ech_t(f) := \deg_{\leq t}(f(\vecx + \vecz) - f(\vecz))$, where $\deg_{\leq t}$ restricts to just those monomials in $\vecx$ of degree at most $t$. 
It is also worth noting that $\Ech_t(f)$ does not have a constant term and this would become useful in the criterion.
	
The operator $\Ech_t(f)$ can be thought of as a vector over the field $\F(\vecz)$ whose coordinates are indexed by monomials $\vecx^\vece$ of degree at most $t$. 
The entry in the coordinate $\vecx^{\vece}$ of $\Ech_t(\vecf)$ is the corresponding \emph{Hasse derivative} of $f$ evaluated at $\vecz$:
\[
	\frac{\abs{\vece}!}{e_1!e_2!\cdots e_n!} \cdot \inparen{\frac{\partial^{\abs{\vece}}f}{\partial x_1^{e_1} \cdots \partial x_n^{e_n}}}(\vecz).
\]
	
The operator $\Ech_t$ however, as defined above, is indexed by $t$. 
Pandey \etal~\cite{PSS18} show that the correct value of $t$ to work with is the \emph{inseparable degree} of the given set of polynomials.	
Formally, we have the following statement.
	
\PSS*
	
We note that at least one direction of this theorem can be slightly generalised to give the following lemma. 
A proof is given here for the sake of completeness, but we note that the steps are almost identical to those in \cite{PSS18}.
	
\begin{restatable}{lemma}{PSSDep}\label{lem:Dep}
	Let $\F$ be an algebraically closed field and $\mathbb{K}$ be an extension field of $\F$. 
	Further, suppose $\set{g_1,\ldots, g_k}$ is a set of $n$-variate polynomials in $\mathbb{K}[\vecy]$ that are $\F$-algebraically dependent. 
	Also, for a generic point $\vecv$, let $\Ech_t(g_i) = \deg_{\leq t}(g_i(\vecy + \vecv) - g_i(\vecv))$. 
	Then for any positive integer $t$, there exists $(\alpha_1, \ldots, \alpha_k) \in \F(\vecg(\vecv))^k \setminus \set{\mathbf{0}}$ such that
	\[
		\sum_{i=1}^k \alpha_i \Ech_t(g_i) \equiv 0 \mod\inangle{\Ech_t(g_1), \ldots, \Ech_t(g_k)}^{\geq 2}_{\F(\vecg(\vecv))} + \inangle{\vecy}^{t+1}
	\]
\end{restatable}

\begin{proof}
	Suppose $\set{g_1, \ldots, g_k}$ are $\F$-algebraically dependent. 
	Then by standard properties of transcendence bases \cite[Theorem 7.20 and 7.18]{K07}, we have that there is an $\F$-algebraically independent subset of $\set{g_1, \ldots, g_k}$, of size $r < k$, that forms a separable transcendence basis.
	Without loss of generality, let that subset be $\set{g_1, \ldots, g_r}$.
		
	Let $A \in \F[u_0, u_1, \ldots, u_r]$ be the minimal annihilating polynomial for $\vecg = \set{g_0, g_1, \ldots, g_r}$ where $g_0:=g_{r+1}$.
	Now since $A(\vecg) = 0$, for formal variables $\vecv$, we have $A(\vecg(\vecy + \vecv)) = 0$.
	Also, from the definition of $\Ech_t(g)$, we have that $g_j(\vecy + \vecv) = g_j(\vecv) + \Ech_t(g_j) \mod \inangle{\vecy}^{t+1}$ for any $j = 0,\ldots, r$. Hence,
	\[
		A(g_0(\vecv) + \Ech_t(g_0),\ldots, g_r(\vecv) + \Ech_t(g_r)) = 0 \mod{\inangle{\vecy}^{t+1}}.
	\]
	
	Using Taylor expansion, we get
	\begin{align*}
		A(g_0(\vecv) + \Ech_t(g_0),\ldots, g_r(\vecv) + \Ech_t(g_r)) & = \sum_{\vece\geq 0} \inparen{\partial_{\vecu^\vece} A}_{\vecu = \vecg(\vecv)} \cdot (\Ech_t(\vecg))^\vece\\
		& =  A(\vecg(\vecv)) + \sum_{i=0}^r \inparen{\partial_{u_i}A}_{\vecu = \vecg(\vecv)} \Ech_t(g_i) \\
		& \quad\quad\quad\mod \inangle{\Ech_t(g_0), \ldots, \Ech_t(g_r)}^{\geq 2}_{\F(\vecg(\vecv))} + \inangle{\vecy}^{t+1}
	\end{align*}
	where the last equality crucially used the fact that the coefficients of $A$ are from $\F$ and hence the linear combinations of $\inangle{\Ech_t(\vecg)}^{\geq 2}$ are over $\F(\vecg(\vecv))$. 
		
	Observe that $A(\vecg(\vecv)) = 0$. Furthermore, since $\set{g_1, \ldots, g_r}$ forms a separable basis, we have that $\partial_{u_0} A$ is a nonzero polynomial. Hence $\partial_{u_0}(A(\vecg(\vecv))) \neq 0$, as $A$ is the minimal degree annihilator for $\vecg$. Therefore, we have a nonzero vector $(\alpha_1,\ldots,\alpha_k) \in \inparen{\F(\vecg(\vecv))}^k$ such that
	\[
		\sum_{i=1}^k \alpha_i \Ech_t(g_i) \equiv 0 \mod\inangle{\Ech_t(g_1), \ldots, \Ech_t(g_k)}^{\geq 2}_{\F(\vecg(\vecv))} + \inangle{\vecy}^{t+1}\qedhere
	\]
\end{proof}
	
\subsubsection*{A different perspective on the criterion}
	
Let $\mathcal{U}_t(\vecf) = \mathcal{U}_t(f_1, \ldots, f_k)$ denote the subspace $\genset{\Ech_t(\vecf)}^{\geq 2}_{\F(\vecz)}$\\ $= \genset{\Ech_t(f_1), \ldots, \Ech_t(f_k)}^{\geq 2}_{\F(\vecz)} \bmod{\genset{\vecx}^{t+1}}$. 
Then, for any $h \in \mathcal{U}_t(\vecf)$, we define the modified Jacobian matrix as follows.
\[
	\PSSJac_t(\vecf, h) = \insquare{\begin{array}{c}  \Ech_t(f_1) + h \\ \Ech_t(f_2) \\ \vdots \\ \Ech_t(f_k) \end{array}}. 
\]
The columns of this matrix are indexed by monomials in $\vecx$ and entries in the column indexed by $\vecx^\vece$ are the coefficient of $\vecx^\vece$ in the corresponding rows.
	
An alternative statement for the PSS criterion is thus, the following.
\begin{theorem}[Alternate Statement for the PSS-criterion] \label{thm:AltPSScriterion}
	Let $\set{f_1, \ldots, f_k}$ be a set of $n$-variate polynomials over a field $\F$ with inseparable degree $t$. 
	Then, they are algebraically independent if and only if for every $h \in \mathcal{U}_t(\vecf)$, $\PSSJac_t(\vecf, h)$ is full rank.
\end{theorem}	
	
We note that \autoref{lem:Dep} can also be viewed from a similar perspective. 
Let $\mathcal{V}_t(g_1, \ldots, g_k)$ denote the subspace $\genset{\Ech_t(g_1), \ldots, \Ech_t(g_k)}^{\geq 2}_{\F(\vecg(\vecv))} \bmod{\genset{\vecy}^{t+1}}$. 
An alternate statement for the lemma is then the following.
	
\begin{lemma}[Alternate statement for \autoref{lem:Dep}]\label{lem:AltDep}
	Let $\F$ be any field and $\mathbb{K}$ be an extension field of $\F$. 
	If $\set{g_1,\ldots, g_k}$ is a set of $n$-variate polynomials in $\mathbb{K}[\vecy]$ that are $\F$-algebraically dependent, then for any positive integer $t$, there exists $h' \in \mathcal{V}_t(g_1, \ldots, g_k)$ such that $\PSSJac_t(\vecg, h')$ is not full rank.
\end{lemma}
	
\section{Rank Condensers from Isolating Weight Assignments}\label{sec:rank-conc}
	
In this section, we focus on \emph{rank-preserving} properties of certain types of matrices. 
These are slight generalisations of similar properties of Vandermonde matrices that were proved by Gabizon and Raz~\cite{GR08} that would be necessary for the application to constructing faithful homomorphisms. 
	
\begin{lemma}\label{lem:vandermonde-with-zeros}
	Suppose we have a matrix of the form:
	\[
		V = \begin{bmatrix}
			s^{w_1} & s^{2w_1} & \ldots & s^{nw_1} \\
			s^{w_2} & s^{2w_2} & \ldots & s^{nw_2} \\
			& & \vdots\\
			s^{w_n} & s^{2w_n} & \ldots & s^{nw_n}
		\end{bmatrix}
	\]
	where $w_i < w_j$ whenever $i < j$. 
	If $V'$ is a matrix obtained from $V$ by replacing some of the non-diagonal entries by zero, then $\det(V') \neq 0$ and furthermore $\deg(\det(V')) = \sum_{i=1}^{n} i \cdot w_i$.
\end{lemma}

\begin{proof}
	Since 
	\[
		\det(V') = \sum_{\sigma \in S_n} \sgn(\sigma) \inparen{\prod_{i \in [n]} V'[i, \sigma(i)]},
	\]
	the monomial corresponding to $\sigma$ being the identity permutation contributes a nonzero monomial of degree $\sum i\cdot w_i$. 
	We will show that all other terms of $\det(V')$ will have smaller degree. 
		
	Suppose $\sigma$ is not the identity permutation, we must have $i \neq \sigma(i)$ for some index $i$; let $i_0$ be the first such index. 
	Define $j$ such that $\sigma(j) = i_0$ and $\pi = \sigma \circ (i_0 \text{ } j)$. 
	Note that $\pi(i_0) = \sigma(j) = i_0$ and fixes the first $i_0$ indices. 
	Furthermore, $\pi(i)= \sigma(i)$ for all $i \neq i_0,j$. Thus,
	\begin{align*}
		\sum_{i=1}^{n} (\pi(i) - \sigma(i)) \cdot w_i &= (\pi(i_0) - \sigma(i_0)) \cdot w_{i_0} + (\pi(j) - \sigma(j)) \cdot w_j \\
		&= (\sigma(j) - \sigma(i_0)) \cdot w_{i_0} + (\sigma(i_0) - \sigma(j)) \cdot w_j \\
		&= (\sigma(i_0) - \sigma(j)) \cdot (w_j - w_{i_0}) > 0
	\end{align*}
	Repeating this exercise until we reach the identity permutation, we have that the monomial contributed by the diagonal has the largest degree. 
\end{proof}
	
\begin{lemma}\label{lem:tranfer-matrix-iwa-rank-extractor}
	Let $A$ be a matrix over a field $\F$ with $k$ rows and columns indexed by monomials in $\vecx$ of degree at most $D$ that is full-rank. 
	Further, let $w = (w_1, \ldots, w_n)$ be an isolating weight assignment for the set of degree $D$ monomials, and let $\wt(\vecx^\vece) = \sum_{i=1}^{n} w_i e_i$.
		
	Suppose $M_\Phi$ is a matrix whose rows are indexed by monomials in $\vecx$ of degree at most $D$, and columns indexed by \emph{pure monomials} $\setdef{y_i^d}{i\in \set{1,\ldots, k}\;,\;d\leq D}$ given by
	\[
		M_\Phi(\vecx^\vece, y_i^d) = \begin{cases}
		s^{i \cdot \wt(\vecx^\vece)} & \text{if $\deg(\vecx^{\vece}) = d$}\\
		0 & \text{otherwise}
		\end{cases}.
	\]
	where $s$ is a formal variable. 
	Then, $\rank_{\F(s)}(A \cdot M_\Phi) = \rank_{\F}(A)$.
\end{lemma}
	
\begin{proof}
	By the Cauchy-Binet formula, if we restrict $M_\Phi$ to a set $T$ of $k$-columns, then
	\[
		\det( A \cdot M_\Phi[T]) = \sum_{\substack{S \subseteq \mathrm{Columns}(A)\\|S| = k}}\det(A[S]) \cdot \det(M_\Phi[S,T])
	\]
	We wish to show that the above sum is nonzero for some choice of columns $T$. 
	We do that by first defining a weight function on minors of $A$, then proving that there is a unique nonzero minor of $A$ of largest weight, and then choosing a set of columns $T$ such that the degree of $\det(M_\Phi[S,T])$ coincides with this chosen weight function.
	Define the \emph{weight} of a minor of $A$ as follows: 
		
	\begin{quote}
		Suppose the columns of the minor is indexed by $S = \set{\vecx^{\mathbf{e_1}}, \ldots, \vecx^{\mathbf{e_k}}}$ with the property that $\wt(\vecx^{\mathbf{e_1}}) < \wt(\vecx^{\mathbf{e_2}}) < \cdots < \wt(\vecx^{\mathbf{e_{k}}})$. 
		Define the weight of this minor as
		\[
			\wt(S) = \sum_{i=1}^{k} i \cdot \wt(\vecx^{\mathbf{e_i}})
		\]
		where, recall, $\wt(\vecx^{\mathbf{e_i}}) = \sum_j w_j \cdot \mathbf{e_i}(j)$. 
	\end{quote}  
		
	\begin{claim} \label{clm:unique-minor-max-wt}
		There is a unique nonzero $k\times k$ minor of $A$ of maximum weight. 
	\end{claim}
	
	\begin{proof}
		Suppose $S_1$ and $S_2$ are two different minors of $A$ with the same weight. 
		We will just identify $S_1$ and $S_2$ by the set of column indices for simplicity. 
		Say $S_1$ has columns indexed by $\vecx^{\mathbf{e_1}}, \ldots, \vecx^{\mathbf{e_{k}}}$ with $\wt(\vecx^{\mathbf{e_1}}) < \wt(\vecx^{\mathbf{e_2}}) < \cdots < \wt(\vecx^{\mathbf{e_{k}}})$ and $S_2$ has columns indexed by $\vecx^{\mathbf{e'_1}}, \ldots, \vecx^{\mathbf{e'_{k}}}$ with $\wt(\vecx^{\mathbf{e'_1}}) < \wt(\vecx^{\mathbf{e'_2}}) < \cdots < \wt(\vecx^{\mathbf{e'_{k}}})$.
			
		Suppose $S_1$ and $S_2$ agree on the first $i$ columns, that is $\vece_j = \vece_j'$ for all $j \leq i$, and say $\wt(\vece_{i+1}) < \wt(\vece_{i+1}')$. 
		By the matroid property, there must be some column $\vecx^{\vece_j'}$ from $S_2$ that we can add to $S_1 \setminus \set{\vecx^{\vece_{i+1}}}$ so that $S = S_1 \setminus \set{\vecx^{\vece_{i+1}}} \union \set{\vecx^{\vece_j'}}$ is also a nonzero minor of $A$. 
		Suppose that
		\[
			\wt(\vecx^{\vece_1}) < \cdots < \wt(\vecx^{\vece_{i+r}}) < \wt(\vecx^{\vece_j'}) < \wt(\vecx^{\vece_{i+r+1}}) < \cdots < \wt(\vecx^{\vece_{k}}).
		\]
		Then,
		\begin{align*}
			\wt(S) & =  \sum_{a=1}^i a \cdot \wt(\vecx^{\vece_a}) + \sum_{a=i+2}^{i+r} (a-1) \cdot \wt(\vecx^{\vece_a}) + (i+r) \wt(\vecx^{\vece_j'}) + \sum_{a=i+r+1}^k a \cdot \wt(\vecx^{\vece_a})\\
			& > \sum_{a=1}^{i} a \cdot \wt(\vecx^{\vece_a}) + (i+1) \wt(\vecx^{\vece_j'}) + \sum_{a=i+2}^{k} a \cdot \wt(\vecx^{\vece_a})\\
			&  > \sum_{a=1}^k a \cdot \wt(\vecx^{\vece_a}) = \wt(S_1)
		\end{align*}
			
		Hence, there cannot be two different nonzero minors of $A$ of the same weight. 
		Thus, the nonzero minor of largest weight is unique. 
	\end{proof}
		
	We will now choose $k$ columns from $M_\Phi$, as follows, in such a way that the degree of the corresponding determinant agrees with the weight function. 
	Note that the matrix $M_\Phi$ has a natural block-diagonal structure based on the degree of the monomials indexing the rows and columns. 
	\begin{itemize}
		\item Let $S_0$ be the unique $k \times k$ minor of $A$ having maximum weight. 
		Further, assume its columns are indexed by $\vecx^{\mathbf{e_1}}, \ldots, \vecx^{\mathbf{e_k}}$ with $\wt(\vecx^{\mathbf{e_1}}) < \wt(\vecx^{\mathbf{e_2}}) < \ldots < \wt(\vecx^{\mathbf{e_k}})$.
		Let $d_i = \deg(\vecx^{\vece_i}) = \sum_j (\vece_i)_j$. 
		\item Choose the columns $T = \set{y_{1}^{d_1}, y_2^{d_2},\dots, y_k^{d_k}}$ of the matrix $M_\Phi$. 
	\end{itemize}
	By \autoref{lem:vandermonde-with-zeros}, for any set of $S'\subseteq \operatorname{Columns}(A)$, we have $\deg(\det(M_\Phi[S',T])) \leq \wt(S')$ and furthermore we also have $\deg(M_\Phi[S_0,T]) = \wt(S_0)$ as we chose the columns $T$ to ensure that the main diagonal of the sub-matrix has only nonzero elements. 
	Hence,
	\[
		\det( A \cdot M_\Phi[T]) = \sum_{\substack{S \subseteq \mathrm{Columns}(A)\\|S| = k}}\det(A[S]) \cdot \det(M_\Phi[S,T]) \neq 0
	\]
	since the contribution from $\det(A[S_0]) \det(M_\Phi[S_0,T])$ is the unique term of highest degree and so cannot be cancelled.  
\end{proof}
	
\section{Construction of Explicit Faithful Maps}\label{sec:recipe}
	
We will be interested in applying a map $\Phi:\F[\vecx] \rightarrow \F(s)[\vecy]$ and study the transformation of the PSS-Jacobian. 
Since the entries of the PSS-Jacobian involve $\Ech_t(f(\vecx)) = \deg_{\leq t}\inparen{f(\vecx + \vecz) - f(\vecz)}$, we would need to also work with $\Ech_t(g(\vecy))$ where $g(\vecy) = f \circ \Phi$. 
To make it easier to follow, we shall use a different name for the variables in the two cases. Hence,
\[
	\Ech_t(f(\vecx)) := \deg_{\leq t}\inparen{f(\vecx + \vecz) - f(\vecz)} \quad,\quad
	\Ech_t(g(\vecy)) := \deg_{\leq t}\inparen{g(\vecy + \vecv) - g(\vecv)}.
\]
	
\subsection{Recipe for constructing faithful maps}
	
Let $f_1, \ldots, f_m \in \F[x_1, \ldots, x_n]$ be polynomials with $\algrank\set{f_1, \ldots, f_m} = k$ and inseparable degree $t$. 
We will work with linear transformations of the form: 
\begin{align*}
	\Phi:x_i &\mapsto a_i y_0 + \sum_{j=1}^k s^{w_i \cdot j} y_j,\quad\text{for all $i\in [n]$},\\
	\Phi_z:z_i &\mapsto a_i v_0 + \sum_{j=1}^k s^{w_i \cdot j} v_j,\quad\text{for all $i\in [n]$}.
\end{align*}
where all the variables on the RHS are formal variables. 
Further, define $\set{g_1, \ldots, g_m} \in \F[\vecy]$ as $g_i = f_i \circ \Phi$ and $\Ech_t(g_i) = \deg_{\leq t}(g_i(\vecy + \vecv) - g_i(\vecv))$.
	
The main lemma of this section is the following \emph{recipe} for constructing faithful maps. 
	
\begin{lemma}[Recipe for faithful homomorphisms]\label{lem:FaithfulMapRecipe} 
	Let $f_1, \ldots, f_m \in \F[\vecx]$ be polynomials such that their algebraic rank is at most $k$ and suppose the inseparable degree is bounded by a constant $t$. 
	Further,
	\begin{itemize}
		\item suppose $\mathcal{G} = (G_1(\alpha),\ldots, G_n(\alpha)) = (a_1, \ldots, a_n)$ is such that for some $\veca \in \mathcal{G}$, the rank of $\PSSJac_t(\vecf, h)$ is preserved after the substitution $\vecz \to \veca$.
		\item suppose $w:[n]\rightarrow \N$ is an isolating weight assignment for the set of $n$-variate monomials of degree at most $t$. 
	\end{itemize}
		
	\noindent
	Then, the homomorphism $\Phi: \F[x_1,\ldots, x_n] \rightarrow \F(s,\alpha)[y_0,\ldots, y_k]$ defined as 
	\[
		\Phi: x_i \mapsto y_0 G_i(\alpha) + \sum_{j=1}^k y_j\cdot  s^{w(i) j},
	\]
	is an $\F$-faithful homomorphism for the set $\set{f_1, \ldots, f_m}$. 
\end{lemma}
	
As mentioned earlier, the rough proof sketch would be to first write the PSS-Jacobian of the transformed polynomials $\vecg$ in terms of $\vecf$, express that as a suitable matrix product, and use some \emph{rank extractor} properties of the associated matrix, as described in \autoref{sec:rank-conc}. 
The rest of this section will execute this sketch. 
	
\begin{lemma}[Evolution of polynomials under $\Phi$] \label{lem:Ht-before-after-Phi}
	Let $\Phi:\vecx \rightarrow \F(s)[\vecy]$ and $\Phi_z:\vecz\rightarrow \F(s)[\vecv]$ be given as above. 
	Further, for any polynomial $h'(a_1,\ldots, a_m) \in \F(\vecg(\vecv))[\mathbf{\veca}]$, define $h(a_1,\ldots, a_m) \in \F(\vecf(\vecz))[\mathbf{\veca}]$ as follows.
	\begin{quote}
		$\coeff_{\veca^\vece}(h)$ is got by replacing every occurrence of $g_i(\vecv)$ by $f_i(\vecz)$ in $\coeff_{\veca^\vece}(h')$
	\end{quote}  
	Then, 
	\[
		h'(\Ech_t(g_1), \ldots, \Ech_t(g_m)) = \Phi \circ \Phi_z(h(\Ech_t(f_1), \ldots, \Ech_t(f_m))).
	\]
\end{lemma}
	
It is worth noting that the polynomial $h(a_1,\ldots, a_m)$ is independent of $s$, by definition. 
This would be crucial later on in the proof. 
	
\begin{proof}
	Firstly, note that $h$ is well defined. 
	This is because by the definition of $\set{g_1, \ldots, g_m}$, if $\coeff_{\veca^\vece}(h') \in \F(\vecg(\vecv))$ has a nonzero denominator then by replacing the $g_i(\vecv)$s with $f_i(\vecz)$ in it, it will continue to remain nonzero.
		
	The claim now follows essentially from the fact that $\Phi$ is linear and homogeneous in $\vecy$.
	\begin{align*}
		\Ech_t(f \circ \Phi)(\vecy, \vecv) &= \deg_{\leq t}\insquare{(f \circ \Phi)(\vecy + \vecv) - (f\circ \Phi)(\vecv)}\\
		& = \deg_{\leq t}\insquare{f(\Phi(\vecx) + \Phi_z(\vecz)) - f(\Phi_z(\vecz))} & \text{(by linearity in $\vecy$)}\\
		& = \Phi \circ \Phi_z(\Ech_t(f)) & \text{(by homogeneity in $\vecy$)}
	\end{align*}
	and it extends to higher degree terms just from the fact that $\Phi$ and $\Phi_z$ are homomorphisms and that $\Phi$ does not change the degree (in $\vecx$ and $\vecy$).
	Further, note that if $h(a_1,\ldots, a_m) = \sum_{\vece} h_\vece \cdot \veca^\vece$ then 
	\[
		h' = \sum_{\vece} \Phi_z(h_\vece) \cdot \veca^\vece.
	\]
	Thus, 
	\begin{align*}
		h'(\Ech_t(g_1), \ldots, \Ech_t(g_m)) &= \sum_{\vece}  \Phi_z(h_\vece) \cdot (\Ech_t(\vecf \circ \Phi))^\vece\\
		&= \sum_{\vece} \Phi_z(h_\vece) \cdot \Phi \circ \Phi_z(\Ech_t(\vecf)^\vece)\\
		& = \sum_{\vece} (\Phi \circ \Phi_z(h_\vece)) \cdot \Phi \circ \Phi_z(\Ech_t(\vecf)^\vece) \quad \quad \text{($h_\vece$ is independent of $\vecx$)}\\
		&= \Phi \circ \Phi_z \inparen{\sum_{\vece} h_\vece \cdot \Ech_t(\vecf)^\vece} \quad \quad \text{($\Phi$ and $\Phi_z$ are homomorphisms)}\\
		&= \Phi \circ \Phi_z(h(\Ech_t(f_1), \ldots, \Ech_t(f_m))) & \qedhere
	\end{align*}
\end{proof}
	
\begin{corollary}[Matrix representation of the evolution] \label{lem:Phi_as_matrix_product}
	Suppose $A'$ is a matrix whose columns are indexed by monomials in $\vecy$. 
	Further suppose a row in $A'$ corresponds to a polynomial, say $h'(\Ech_t(\vecg)) = h'(\Ech_t(g_1), \ldots, \Ech_t(g_m)) \in \F(\vecg(\vecv))[\vecy]$, whose entry in the column indexed by $\vecy^\vece$ is $\coeff_{\vecy^\vece}(h'(\Ech_t(\vecg))) \in \F(\vecv, \vecs)$. 
	If $A$ is the corresponding matrix (having entries from $\F(\vecz)$) with columns indexed by monomials in $\vecx$ and the corresponding row being $h(\Ech_t(f_1), \ldots, \Ech_t(f_m)) \in \F(\vecf(\vecz))[\vecx]$ as described in \autoref{lem:Ht-before-after-Phi}, then
	\[
		A' = \Phi_z(A) \times \widetilde{M_\Phi}
	\]
	where $\widetilde{M_\Phi}(\vecx^\vece, \vecy^\vecd) = \coeff_{\vecy^\vecd}(\Phi(\vecx^{\vece}))$.
\end{corollary}
	
\begin{proof} 
	Suppose $h(\Ech_t(f_1), \ldots, \Ech_t(f_m)) = \sum_{\vece} h_\vece(\vecz) \cdot \vecx^{\vece}$. 
	Then,
	\begin{align*}
		h'(\Ech_t(g_1), \ldots, \Ech_t(g_m)) & = \Phi \circ \Phi_z(h(\Ech_t(f_1), \ldots, \Ech_t(f_m))) & \text{(by \autoref{lem:Ht-before-after-Phi})}\\
		& = \sum_{\vece} h_\vece(\Phi_z(\vecz)) \cdot \Phi(\vecx^{\vece}) \\
		& = \sum_{\vece} h_\vece(\Phi_z(\vecz))\cdot \inparen{\sum_{\vecd} \coeff_{\vecy^\vecd}(\Phi(\vecx^{\vece})) \cdot \vecy^\vecd}\\
		& = \sum_{\vecd} \inparen{\sum_{\vece} h_\vece(\Phi_z(\vecz)) \cdot \coeff_{\vecy^\vecd}(\Phi(\vecx^{\vece}))} \cdot \vecy^\vecd
	\end{align*}
	
	Thus, the coefficient of $\vecy^\vecd$ in $h'(\Ech_t(g_1), \ldots, \Ech_t(g_m))$ is
	\[
		\sum_{\vece} \Phi_z(h_\vece(\vecz)) \cdot \coeff_{\vecy^\vecd}(\Phi(\vecx^{\vece}))
	\]
	which gives the required matrix decomposition.
\end{proof}
	
We are now in a position to prove \autoref{lem:FaithfulMapRecipe}. 
	
\begin{proof}[Proof of \autoref{lem:FaithfulMapRecipe}]
	Without loss of generality, say $\set{f_1,\ldots, f_k}$ is an algebraically independent set. 
	We wish to show that if $g_i = f_i \circ \Phi$, then $\set{g_1, \ldots, g_k}$ is an $\F$-algebraically independent set as well. 
	Assume on the contrary that $\set{g_1, \ldots, g_k}$ is an $\F$-algebraically dependent set. 
	Then for $t$ being the inseparable degree of $\set{f_1, \ldots, f_k}$, by \autoref{lem:AltDep}, there exists
	\[
		h' \in \mathcal{V}_t(g_1, \ldots, g_k) := \genset{\Ech_t(g_1), \ldots, \Ech_t(g_k)}^{\geq 2}_{\F(\vecg(\vecv))} \bmod{\genset{\vecy}^{t+1}}
	\]
	such that $\PSSJac_t(\vecg, h')$ is not full rank. 
	Without loss of generality, we can assume that the entries of $\PSSJac_t(\vecg, h')$ are denominator-free by clearing out any denominators. 
	Corresponding to $h'$, define $h$ as in \autoref{lem:Ht-before-after-Phi}, which would also satisfy that
	\[
		h \in \mathcal{U}_t(f_1, \ldots, f_k) := \genset{\Ech_t(f_1), \ldots, \Ech_t(f_k)}^{\geq 2}_{\F(\vecz)} \bmod{\genset{\vecx}^{t+1}}.
	\]
	It is worth stressing the fact that the polynomial $h$ is independent of the variable $s$. 
	Then by \autoref{lem:Phi_as_matrix_product} we get
	\[
		\PSSJac_t(\vecg, h') = \Phi_z(\PSSJac_t(\vecf, h)) \times \widetilde{M_\Phi}.
	\]
	Now, if we substitute $v_0 = 1$ and $v_i =0$ for every $i \in [k]$, we get
	\[
		\PSSJac_t(\vecg, h')(v_0 = 1, v_1 = \ldots = v_k = 0) = \PSSJac_t(\vecf, h) (\vecz = \mathbf{G}(\alpha)) \times \widetilde{M_\Phi}.
	\]
	But since $\set{f_1, \ldots, f_k}$ is algebraically independent, \autoref{thm:AltPSScriterion} yields that $\PSSJac_t(\vecf, h)$ has full rank. 
	Thus, for the correct choice of $\alpha$, $\PSSJac_t(\vecf, h) (\vecz = \mathbf{G}(\alpha))$ also has full rank by the property we assumed $\mathcal{G}$ has. 
	Most crucially, the matrix $\PSSJac_t(\vecf,h)$ is independent of the variable $s$.
		
	To complete the proof, we need to show that multiplication by $\widetilde{M_\Phi}$ continues to keep this full rank to contradict the initial assumption that $\PSSJac_t(\vecg,h')$ was not full rank. 
		
	Finally note that for the $\Phi$ we have defined, $\widetilde{M_{\Phi}}$ restricted to only the \emph{pure monomial} columns
	\[
		\setdef{y_i^j}{i \in \set{1,\ldots, k}\;,\;j \in \set{0,1,\ldots, t}},
	\]
	is the same as $M_\Phi$ as defined in \autoref{lem:tranfer-matrix-iwa-rank-extractor}. 
	Further, $w$ is an isolating weight assignment for the set of $n$-variate monomials of degree at most $t$, we satisfy the requirements of  \autoref{lem:tranfer-matrix-iwa-rank-extractor}. 
	Hence, by \autoref{lem:tranfer-matrix-iwa-rank-extractor},
	\begin{align*}
		\rank_{\F(s,\alpha)}\inparen{\PSSJac_t(\vecg,h')(v_0 = 1, v_1 = \ldots, v_k = 0)} & = \rank_{\F(\alpha)}\PSSJac_t(\vecf, h)(\vecz = \mathbf{G}(\alpha))\\
		\implies \rank_{\F(s,\alpha,\vecv)}\inparen{\PSSJac_t(\vecg,h')}   & \geq \rank_{\F(\alpha)}\PSSJac_t(\vecf, h)(\vecz = \mathbf{G}(\alpha))\\
		& = k,
	\end{align*}
	which contradicts our assumption that it was not full rank. 
	Hence, it must indeed be the case that $\set{f_1 \circ \Phi,\ldots, f_k \circ \Phi}$ is $\F$ - algebraically independent.
\end{proof} 
	
\section{Explicit faithful maps and PIT applications in restricted settings}\label{sec:execution}
	
We now describe some specific instantiations of the recipe given by \autoref{lem:FaithfulMapRecipe} in restricted settings. 
Let us first recall the statement of the main theorem. 
	
\FaithfulMaps*
	
\begin{proof}
	By \autoref{lem:FaithfulMapRecipe}, $\Phi: \F[x_1,\ldots, x_n] \rightarrow \F(s,\alpha)[y_0,\ldots, y_k]$ defined as 
	\[
		\Phi: x_i \mapsto y_0 G_i(\alpha) + \sum_{j=1}^k y_j\cdot  s^{w(i) j},
	\]
	is a faithful homomorphism for the set $\set{f_1, \ldots, f_m}$ if $w = (w_1, \ldots, w_n)$ is an isolating weight assignment for $n$-variate monomials of degree at most $t$, and for any $h \in \mathcal{U}_t(\vecf)$, $\mathcal{G} = (G_1(\alpha),\ldots, G_n(\alpha))$ is such that the rank of $\PSSJac_t(\vecf, h)$ is preserved after the substitution $\vecz \to \veca$ for some $\veca \in \mathcal{G}$. 
	We define the weight using the standard hashing techniques \cite{KS01,AB03}.
		
	\paragraph*{Defining $w$} Define $w : [n] \to \N$ as 
	\[
		w(i) = (t+1)^i \pmod{p}
	\]
	where $t$ is the inseparable degree. 
	
	Assuming $t$ to be a constant, there are only $\poly(n)$ many distinct monomials in $\vecx$ of degree at most $t$. 
	Thus, standard results by Klivans and Spielman~\cite{KS01} or Agrawal and Biswas~\cite{AB03} shows that it suffices to go over $\poly(n)$ many `$p$'s before $w$ isolates all monomials in $\vecx$ of degree at most $t$. 

	Let $\PSSJac_t(\vecf)$ be the matrix with columns indexed by monomials in $\vecx$ of degree at most $t$ and rows by $k$-variate monomials $\veca^\vece$ in degree at most $t$, defined as follows.
	\[
		\PSSJac_t(\vecf)[\veca^\vece, \vecx^\vecd] = \coeff_{\vecx^\vecd}(\Ech_t(\vecf)^\vece)
	\]
	Set $K = {k+t \choose t}$ to be the number of rows in $\PSSJac_t(\vecf)$. 
	Then the following is true.
		
	\begin{claim}
		If $\mathcal{G}$ is a hitting set generator for every $K' \times K'$ minor of $\PSSJac_t(\vecf)$ where $K' \leq K$, then the rank of $\PSSJac_t(\vecf, h)$ is preserved for every $h \in \mathcal{U}_t(\vecf)$.
	\end{claim}
	
	\begin{proof}
		We need to show that there is an $\veca$ in $\mathcal{G}$ which has the following property:
			
		\begin{quote}
			For any $h \in \mathcal{U}_t(\vecf)$, if $\set{\Ech_t(f_1) + h, \Ech_t(f_2), \ldots, \Ech_t(f_k)}$ are linearly independent, then so are $\set{\Ech_t(f_1)(\veca) + h(\veca), \Ech_t(f_2)(\veca), \ldots, \Ech_t(f_k)(\veca)}$.
		\end{quote} 
			
		Now suppose this is not the case. 
		Then it must be the case that without loss of generality, some $h \in \mathcal{U}_t(\vecf)$, $\PSSJac_t(\vecf, h)$ has full rank but for any $\veca \in \mathcal{G}$,  
		\[
			\alpha_1 (\Ech_t(f_1)(\veca) + h(\veca)) + \sum_{i=2}^k (\alpha_i \cdot \Ech_t(f_i)(\veca)) = 0.
		\]
		Here, not all of $\set{\alpha_i}_{i \in [k]}$ are zero. 
		However by our hypothesis, this would mean that
		\[
			\alpha_1 (\Ech_t(f_1) + h) + \sum_{i=2}^k (\alpha_i \cdot \Ech_t(f_i)) \neq 0.
		\]
			
		Let $\mathcal{B}$ be a basis of the rows in $\PSSJac_t(\vecf, h)$. 
		Then each of $\set{\Ech_t(f_1) + h, \Ech_t(f_2), \ldots, \Ech_t(f_k)}$ can be written in terms of rows in $\mathcal{B}$. 
		Thus, the above statement can be rewritten as
		\[
			\sum_{i=1}^{K'} \beta_i \cdot b_i = \alpha_1 (\Ech_t(f_1)+h) + \sum_{i=2}^k (\alpha_i \cdot \Ech_t(f_i)) \neq 0
		\]
		where $\set{\beta_i}_{i \in [K']}$ are some scalars, $b_i \in \mathcal{B}$ and $K' = \abs{\mathcal{B}}$.
			
		This shows that not all $\set{\beta_i}_{i=1}^{K'}$ can be zero. 
		Now since $\mathcal{G}$ is a hitting set generator for every $K' \times K'$ minor in $\PSSJac_t(\vecf)$, there is some $\veca \in \mathcal{G}$ such that $\set{b_i(\veca)}_{i \in [K']}$ continue to remain linearly independent. 
		Thus, $\sum_{i=1}^{K'} \beta_i \times b_i(\veca)\neq 0$, since not all $\set{\beta_i}_{i \in [K']}$ is zero. 
		However, this shows that
		\[
			\alpha_1 (\Ech_t(f_1)(\veca)+h(\veca)) + \sum_{i=2}^k (\alpha_i \cdot \Ech_t(f_i)(\veca)) = \sum_{i=1}^{K'} \beta_i \times b_i(\veca) \neq 0.
		\]
			
		This contradicts our assumption, and so it must be the case that for any $h \in \mathcal{U}_t(\vecf)$, the rank of $\PSSJac_t(\vecf, h)$ is preserved.
	\end{proof}
		
	Now it is only a question of finding a hitting set generator of low degree, for every $K' \times K'$ minor of $\PSSJac_t(\vecf)$ where $K' \leq K$.
		
	\subsection*{Defining $\mathcal{G}$ when $f_i$'s are sparse} 
	When the $f_i$'s are $s$-sparse, every entry of $\PSSJac(\vecf)$ is a sum of products of at most $t$ Hasse-derivatives of the $f_i$'s. 
	Further the number of such products is at most $\binom{n+t}{t}$, and hence each entry of $\PSSJac(\vecf)$ has sparsity at most $\binom{n+t}{t}\cdot s^t$. 
	When $k,t$ are constants, then any $K \times K$ minor of $\PSSJac(\vecf)$ has sparsity $s^{O(1)}$ and hence standard hitting-set generators for sparse polynomials \cite{KS01,AB03} would be sufficient in this setting.
		
	\subsection*{Defining $\mathcal{G}$ when $f_i$s are products of variable disjoint, multilinear, sparse polynomials}
	In exactly along the same lines as Agrawal \etal~\cite{ASSS16}, we can construct hitting-set generators for minors of $\PSSJac(\vecf)$ when each $f_i$ is a product of variable disjoint, multilinear, sparse polynomials.
		
	The key observation is that when $k,t = O(1)$, any $K\times K$ minor of $\PSSJac(\vecf)$ only involves derivatives over constantly many variables, say $x_1,\dots, x_\ell$ with $\ell \leq Kt$. Since each $f_i$ is a product of variable disjoint sparse polynomials, each row of this submatrix can be expressed as a common factor $F$ and a product of $\ell$ sparse polynomials. 
	The reason is as follows.
	\begin{quote}
		If $f = g . g'$ where $g'$ is independent of variables in $S \subseteq \set{x_1, \ldots, x_n}$, then for any monomial $\vecx^\vece$ that depends only on $S$ we have 
		\[
			\coeff_{\vecx^\vece}(\Ech_t(f)) = \coeff_{\vecx^\vece}(\Ech_t(g)) . g'(z).
		\]
	\end{quote} 
	Hence, the determinant of this matrix is a product of sparse polynomials (each of sparsity at most $s^{Kt} = \poly(s)$ when $k,t = O(1)$). 
	Once again, standard hitting-set generators for sparse polynomials \cite{KS01,AB03} are sufficient in this case as well.
\end{proof}
	
\subsection{Applications to PIT}
	
Using \autoref{lem: connect}, two straightforward corollaries for PIT for related models. 
	
\CorSparsePIT*
	
\begin{proof}
	Without loss of generality, we may assume that $\F$ is algebraically closed (since nonzeroness of polynomials remain unchanged when interpreted as polynomials over an extension). 
	Suppose $\set{f_1,\ldots, f_k}$ is a separable transcendence basis for $\set{f_1,\ldots, f_m}$ with inseparable degree $t$.
		
	By \autoref{thm:FaithfulConstruct}, we have a polynomial sized list of maps $\set{\Phi_i: \F[\vecx] \rightarrow \F[s,y_0,\ldots, y_k,\alpha]}$, each of degree $\poly(n)$ such that at least one of them is $\F$-faithful for $\set{f_1,\ldots, f_k}$ (and hence also for $\set{f_1,\ldots, f_m}$); let $\Phi$ be such a $\F$-faithful homomorphism. From the construction of \autoref{thm:FaithfulConstruct}, the homomorphism $\Phi$ has degree $\poly(s')$.  
	By \autoref{lem: connect}, we know that $C(f_1,\ldots, f_m) = 0$ if and only if $\Phi(C(f_1,\ldots, f_m))$ is zero. Now that $\Phi(C(f_1,\ldots, f_m))$ is a polynomial in $k+3 = O(1)$ variables, we can use the hitting set obtained from the polynomial identity lemma \cite{O22,DL78,S80,Z79} to give hitting set of size $\poly(s',\deg(C))$ for $C(f_1,\ldots, f_m)$. 
\end{proof}
	
Along exactly the same lines, we get the following corollary in the case when we are working with depth-$4$ multilinear circuits of small algebraic rank and inseparable degree. 
	
\CorMLPIT*
	
As mentioned in the introduction, the above result is incomparable with the PIT results of Pandey \etal~\cite{PSS18} and Kumar and Saraf~\cite{KS17}. 
	
\section{Conclusion and open problems}
	
We studied the task of constructing faithful homomorphisms in the finite characteristic setting and extended the results of Agrawal \etal~\cite{ASSS16} in the setting when the inseparable degree is bounded. There are some very natural open problems in this context.
	
\begin{itemize}
	\item Are the homomorphisms constructed in the paper also $\F(s)$-faithful homomorphisms?
		
	Our proof only provides a recipe towards constructing $\F$-faithful homomorphisms due to technical obstacles involving the criterion for algebraic independence over finite characteristic fields. 
	The exact point where it fails is in the proof of \autoref{lem:FaithfulMapRecipe}. 
	It is crucial that $h \in \mathcal{U}_t(\vecf)$ is $s$-free for our proof to work. This is not an issue in characteristic zero fields and Agrawal \etal~\cite{ASSS16} construct $\F(s)$-faithful homomorphisms.
		
	\item How crucial is the notion of inseparable degree in the context of testing algebraic independence?
		
	The criterion of Pandey, Saxena and Sinhababu~\cite{PSS18} crucially depends on this field theoretic notion and there seems to be compelling algebraic reasons to believe that this is necessary. 
	However, as mentioned earlier, Guo, Saxena and Sinhababu~\cite{GSS19} showed that algebraic independence testing is in $\AM \cap \coAM$ and this proof has absolutely no dependence on the inseparable degree.
\end{itemize}

\section*{Acknowledgements}
We acknowledge support of the Department of Atomic Energy, Government  of India, under project number RTI4001.
	
\bibliographystyle{customurlbst/alphaurlpp}
\bibliography{references}
	
\end{document}

%% file: thmmacros.tex
\usepackage{amsthm}
\usepackage{thmtools,thm-restate}

\numberwithin{equation}{section}
\declaretheoremstyle[bodyfont=\it,qed=\qedsymbol]{noproofstyle}

\declaretheorem[name=Observation,numbered=no]{observation*}

\declaretheorem[numberlike=equation]{theorem}
\declaretheorem[numberlike=equation,style=noproofstyle,name=Theorem]{theoremwp}
\declaretheorem[name=Theorem,numbered=no]{theorem*}

\declaretheorem[numberlike=equation]{lemma}
\declaretheorem[name=Lemma,numbered=no]{lemma*}

\declaretheorem[numberlike=equation]{corollary}
\declaretheorem[name=Corollary,numbered=no]{corollary*}

\declaretheorem[name=Proposition,numbered=no]{proposition*}

\declaretheorem[numberlike=equation]{claim}
\declaretheorem[name=Claim,numbered=no]{claim*}

\declaretheorem[name=Conjecture,numbered=no]{conjecture*}

\declaretheorem[name=Question,numbered=no]{question*}

\declaretheoremstyle[bodyfont=\it,qed=$\lozenge$]{defstyle} 

\declaretheorem[numberlike=equation,style=defstyle]{definition}
\declaretheorem[unnumbered,name=Definition,style=defstyle]{definition*}

\declaretheorem[unnumbered,name=Example,style=defstyle]{example*}

\declaretheorem[unnumbered,name=Notation=defstyle]{notation*}

\declaretheorem[unnumbered,name=Construction,style=defstyle]{construction*}

\declaretheorem[numberlike=equation,style=defstyle]{remark}
\declaretheorem[unnumbered,name=Remark,style=defstyle]{remark*}


%% file: lazymacros.tex
\newcommand{\ckt}{\textbf{$\mathcal{C}$}}
\newcommand{\genset}[1]{\inangle{#1}} 

\renewcommand{\phi}{\varphi}
\renewcommand{\epsilon}{\varepsilon}

\newcommand{\J}{\operatorname{\mathbf{J}}}
\newcommand{\PSSJac}{\operatorname{\mathsf{PSSJac}}}
\newcommand{\Ech}{\textbf{$\mathcal{H}$}}
\newcommand{\algrank}{\operatorname{\mathsf{algrank}}}

\newcommand{\wt}{\operatorname{\mathsf{wt}}}
\newcommand{\coeff}{\operatorname{\mathsf{coeff}}}

\newcommand{\insepdeg}{\operatorname{\mathsf{insep-deg}}}
\newcommand{\Sep}{\operatorname{Sep}}


%% file: customurlbst/bibmacros.tex
\usepackage{nth}
\usepackage{intcalc}
\usepackage{xstring}

\usepackage{ifpdf}
\ifpdf
\else
\usepackage[quadpoints=false]{hypdvips}
\fi

\newcommand{\ECCCurl}[2]{http://eccc.hpi-web.de/report/\ifnumcomp{#1}{>}{93}{19}{20}#1/#2/}

\newcommand{\shortECCC}[2]{\texttt{\href{http://eccc.hpi-web.de/report/\ifnumcomp{#1}{>}{93}{19}{20}#1/#2/}{eccc:TR#1-#2}}}

\newcommand{\parseECCC}[1]{
\StrSubstitute{#1}{TR}{}[\tmpstring]%
\IfSubStr{\tmpstring}{/}{ 
\StrBefore{\tmpstring}{/}[\ecccyear]%
\StrBehind{\tmpstring}{/}[\ecccreport]%
}{
\StrBefore{\tmpstring}{-}[\ecccyear]%
\StrBehind{\tmpstring}{-}[\ecccreport]%
}%
\shortECCC{\ecccyear}{\ecccreport}}

%% file: main.bbl
\begin{thebibliography}{ASSS16}

\bibitem[AB03]{AB03}
Manindra Agrawal and Somenath Biswas.
\newblock \href {http://dx.doi.org/10.1145/792538.792540} {Primality and
  identity testing via Chinese remaindering}.
\newblock {\em J. {ACM}}, 50(4):429--443, 2003.

\bibitem[ASSS16]{ASSS16}
Manindra Agrawal, Chandan Saha, Ramprasad Saptharishi, and Nitin Saxena.
\newblock \href {http://dx.doi.org/10.1137/130910725} {Jacobian Hits Circuits:
  Hitting Sets, Lower Bounds for Depth-D Occur-k Formulas and Depth-3
  Transcendence Degree-k Circuits}.
\newblock {\em {SIAM} J. Comput.}, 45(4):1533--1562, 2016.

\bibitem[BMS13]{BMS13}
Malte Beecken, Johannes Mittmann, and Nitin Saxena.
\newblock \href {http://dx.doi.org/10.1016/j.ic.2012.10.004} {Algebraic
  independence and blackbox identity testing}.
\newblock {\em Inf. Comput.}, 222:2--19, 2013.

\bibitem[DL78]{DL78}
Richard~A. DeMillo and Richard~J. Lipton.
\newblock \href {http://dx.doi.org/10.1016/0020-0190(78)90067-4} {{A
  Probabilistic Remark on Algebraic Program Testing}}.
\newblock {\em Information Processing Letters}, 7(4):193--195, 1978.

\bibitem[GR08]{GR08}
Ariel Gabizon and Ran Raz.
\newblock \href {https://doi.org/10.1007/s00493-008-2259-3} {Deterministic
  extractors for affine sources over large fields}.
\newblock {\em Comb.}, 28(4):415--440, 2008.

\bibitem[GSS19]{GSS19}
Zeyu Guo, Nitin Saxena, and Amit Sinhababu.
\newblock \href {http://dx.doi.org/10.4086/toc.2019.v015a016} {Algebraic
  Dependencies and {PSPACE} Algorithms in Approximative Complexity over Any
  Field}.
\newblock {\em Theory Comput.}, 15:1--30, 2019.

\bibitem[Isa94]{I94}
Irving~Martin Isaacs.
\newblock {\em {Character theory of finite groups}}.
\newblock Dover publications Inc., New York, 1994.

\bibitem[Jac41]{J41}
C.G.J. Jacobi.
\newblock \href {http://eudml.org/doc/147138} {De Determinantibus
  functionalibus.}
\newblock {\em Journal für die reine und angewandte Mathematik}, 22:319--359,
  1841.

\bibitem[Kay09]{K09}
Neeraj Kayal.
\newblock \href {http://dx.doi.org/10.1109/CCC.2009.37} {The Complexity of the
  Annihilating Polynomial}.
\newblock In {\em Proceedings of the 24th Annual {IEEE} Conference on
  Computational Complexity, {CCC} 2009, Paris, France, 15-18 July 2009}, pages
  184--193, 2009.

\bibitem[Kna07]{K07}
Anthony~W Knapp.
\newblock {\em Advanced algebra}.
\newblock Springer Science \& Business Media, 2007.

\bibitem[KS01]{KS01}
Adam Klivans and Daniel~A. Spielman.
\newblock \href {http://dx.doi.org/10.1145/380752.380801} {{Randomness
  efficient identity testing of multivariate polynomials}}.
\newblock In {\em \STOC{2001}}, pages 216--223, 2001.

\bibitem[KS17]{KS17}
Mrinal Kumar and Shubhangi Saraf.
\newblock \href {http://dx.doi.org/10.4086/toc.2017.v013a006} {Arithmetic
  Circuits with Locally Low Algebraic Rank}.
\newblock {\em Theory of Computing}, 13(1):1--33, 2017.

\bibitem[Ore22]{O22}
{\O}ystein Ore.
\newblock {\"{U}}ber h{\"{o}}here Kongruenzen.
\newblock {\em Norsk Mat. Forenings Skrifter}, 1(7):15, 1922.

\bibitem[Oxl92]{O92}
James~G. Oxley.
\newblock {\em {Matroid theory}}.
\newblock Oxford University Press, 1992.

\bibitem[PSS18]{PSS18}
Anurag Pandey, Nitin Saxena, and Amit Sinhababu.
\newblock \href {http://dx.doi.org/10.1007/s00037-018-0167-5} {Algebraic
  independence over positive characteristic: New criterion and applications to
  locally low-algebraic-rank circuits}.
\newblock {\em Comput. Complex.}, 27(4):617--670, 2018.

\bibitem[Sch80]{S80}
Jacob~T. Schwartz.
\newblock \href {http://dx.doi.org/10.1145/322217.322225} {{F}ast
  {P}robabilistic {A}lgorithms for {V}erification of {P}olynomial
  {I}dentities}.
\newblock {\em Journal of the ACM}, 27(4):701--717, 1980.

\bibitem[Zip79]{Z79}
Richard Zippel.
\newblock \href {http://dx.doi.org/10.1007/3-540-09519-5_73} {Probabilistic
  algorithms for sparse polynomials}.
\newblock In {\em Symbolic and Algebraic Computation, {EUROSAM} '79, An
  International Symposiumon Symbolic and Algebraic Computation}, volume~72 of
  {\em Lecture Notes in Computer Science}, pages 216--226. Springer, 1979.

\end{thebibliography}
